\documentclass{svmult}

\usepackage{graphicx}        
\usepackage{multicol}        
\usepackage[bottom]{footmisc}
\usepackage{amssymb, bm}
\usepackage{lscape}
\usepackage{color}
\definecolor{lgray}{gray}{.75}
\def\0{\phantom{0}}
\def\.{\phantom{.}}

\makeindex             

\begin{document}

\title*{Molecular Modeling and Simulation of Thermophysical Properties: Application to Pure Substances and Mixtures}
\titlerunning{Molecular Modeling and Simulation}
\author{Bernhard Eckl \and Martin Horsch \and Jadran Vrabec \and Hans Hasse}
\authorrunning{B. Eckl \textit{et al.}}
\institute{Institut f\"ur Technische Thermodynamik und Thermische Verfahrenstechnik, Universit\"at Stuttgart, D-70550 Stuttgart, Germany \texttt{vrabec@itt.uni-stuttgart.de}}
%
%
\maketitle

\section{Introduction}\label{sec_intro}
For development of new processes and optimization of existing plants, the knowledge of reliable thermophysical data is crucial. Classical methods, next to the often expensive and time consuming experimental determination, are based on models of Gibbs excess enthalpy (G$^\mathrm{E}$-models) or equations of state. These phenomenological approaches usually include a large number of adjustable parameters to describe the real behavior quantitatively correct. As in most cases these parameters have no physical meaning, their methodic deduction is rarely possible and a fit to extensive experimental data is unavoidable. Furthermore, extrapolations may be too unreliable for engineering applictions.

Molecular modeling and simulation offers another approach to thermophysical properties on a much better physical basis. First simulations of this type were done by Metropolis \textit{et al.} in 1953 \cite{MRRTT53}. By solely defining the interactions between molecules, it is possible to calculate the relevant thermophysical properties with statistical methods. There are two basic techniques which are fundamentally different: molecular dynamics (MD) and Monte Carlo (MC). The goal of both approaches is to sample the phase space, i.e. all available positions and orientations of the molecules under the given boundaries, and an estimation of derivatives of the partition function. Macroscopic properties like pressure or enthalpy are calculated with statistical thermodynamic methods. Detailed descriptions are given, e.g. by Allen and Tildesley \cite{AT87}.

In the present project, the molecular interactions are described by effective pair potentials. These potentials distinguish dispersive, repulsive, and electrostatic interactions as well as interactions resulting from hydrogen bonding. The dispersive and repulsive part is approximated by Lennard-Jones sites due to their computational efficiency. Electrostatics is modeled by partial charges, ideal point dipoles and ideal point quadrupoles. For hydrogen bonding good results were obtained using eccentric partial charges.


Due to a consistent modeling of the pure substances, an application to mixtures is straightforward. Mixed electrostatic potentials are given directly by the underlying laws of electrostatics. The parameters of the unlike Lennard-Jones interactions are taken from combining rules, where good results are obtained with the Lorentz-Berthelot rule \cite{Lorentz81, Berthelot98}. If needed, one adjustable state-independent parameter can be introduced to fit the binary simulation results to experimental data. No further parameters were to be used for the description of ternary or higher mixtures.

In the present work, different applications of molecular modeling and simulation are presented. Within the project MMSTP, two new molecular models for ethylene oxide and ammonia were developed and subsequently used for prediction of different pure substance properties. Next to this, molecular modeling and simulation was applied on the prediction of nucleation processes. Furthermore, within the present project a comprehensive study on the determination of vapor-liquid equilibria of binary and ternary mixtures is performed. This is a still ongoing task and thus is not included in the present report.

Results of this work are consistently published in peer-reviewed international journals. The following publications contribute to the present project:
\begin{itemize}
\item B. Eckl, J. Vrabec, M. Wendland, and H. Hasse: Thermophysical properties of dry and humid air by molecular simulation - dew point calculations in a new ensemble. In preparation.
\item B. Eckl, J. Vrabec, and H. Hasse: A set of new molecular models based on quantum mechanical \emph{ab initio} calculations. \textit{J. Phys. Chem. B}, in press (2008).
\item M. Horsch, J. Vrabec, and H. Hasse: Molecular dynamics based analysis of nucleation and surface energy of droplets in supersaturated vapors of methane and ethane. \textit{ASME J. Heat Transfer}, in press (2008).
\item M. Horsch, J. Vrabec, and H. Hasse: Modification of the classical nucleation theory based on molecular simulation data for critical nucleus size and nucleation rate. \textit{Phys. Rev. E}, 78: 011603 (2008).
\item M. Horsch, J, Vrabec, M. Bernreuther, S. Grottel, G. Reina, A. Wix, K. Schaber, and H. Hasse: Homogeneous nucleation in supersaturated vapors of methane, ethane, and carbon dioxide predicted by brute force molecular dynamics. \textit{J. Chem. Phys.}, 128: 164510 (2008).
\item B. Eckl, J. Vrabec, and H. Hasse: An optimized molecular model for ammonia. \textit{Mol. Phys.}, 106: 1039 (2008).
\item B. Eckl, J. Vrabec, and H. Hasse: On the Application of Force Fields for Predicting a Wide Variety of Properties: Ethylene Oxide as an Example. \textit{Fluid Phase Equilib.}, in press (2008).
\end{itemize}

The present status report briefly shows results of the pure substances ethylene oxide and ammonia, demonstrating the outstanding predictive power of the applied methods. Secondly, the application to nucleation processes is presented. Next to a pure description, the authors were able to further study the underlying effects on the molecular level and to deduce a modification to classical nucleation theory. Futher details are given in the references mentioned above.

\section{Prediction of a wide variety of properties for ethylene oxide}
The predictive and extrapolative power of molecular modeling and simulation is of particular interest for industrial applications, where a broad variety of properties is needed but often not available. To discuss this issue, the Industrial Simulation Collective \cite{IFPSC} has organized the Fourth Industrial Fluid Properties Simulation Challenge as an international contest. The task in 2007 was to calculate for ethylene oxide on the basis of a single molecular model a total of 17 different properties from three categories. It should be noted that our contribution was awarded with the first price \cite{IFPSC}.

Ethylene oxide (C$_2$H$_4$O) is a widely used intermediate in the chemical industry. In 2006, 18 million metric tons were produced mostly by direct oxidation of ethylene, over 75~\% of which were used for ethylene glycols production. Despite its technical and economical importance, experimental data on thermophysical properties of ethylene oxide are rare, apart from basic properties at standard conditions \cite{DIPPR}. This lack of data is mainly due to the hazardous nature of ethylene oxide. It is highly flammable, reactive, explosive at elevated temperatures, toxic, carcinogenic, and mutagenic. Therefore, it is an excellent example to show that molecular modeling and simulation can serve as a reliable route for obtaining thermophysical data in cases, where avoiding experiments is highly desirable.

In the present work, a new molecular model for ethylene oxide was developed. This model is based on prior work at our institute \cite{Stoll05} and was further optimized to experimental vapor-liquid equilibria (VLE), i.e. saturated liquid density, vapor pressure, and enthalpy of vaporization. Using this model, phase equilibria, thermal, caloric, transport properties, and surface tension were predicted and compared to experimental results, where possible.

\subsection{Optimization of the molecular model for ethylene oxide}\label{sec_eox}
\label{Modell}
The molecular model for ethylene oxide consists of three Lennard-Jones (LJ) sites (one for each methylene group and one for the oxygen atom) plus one static point dipole. Thus, the united-atom approach was used. Due to the fact that ethylene oxide is a small molecule, the internal degrees of freedom may be neglected and the model was assumed to be rigid. Experimental data on the molecular structure was used to specify the geometric location of the LJ sites, cf. \cite{Stoll05}. The dipole was located in the center of mass.
%
%

A set of five adjustable parameters, i.e., the four LJ parameters and the dipole moment, was optimized. This was done by a Newton scheme using correlations to experimental bubble density, vapor pressure, and enthalpy of vaporization data over the full range of the VLE between triple point and critical point.

\begin{figure}[htb]
\begin{center}
\hspace*{-0.8cm}\includegraphics[scale=0.3]{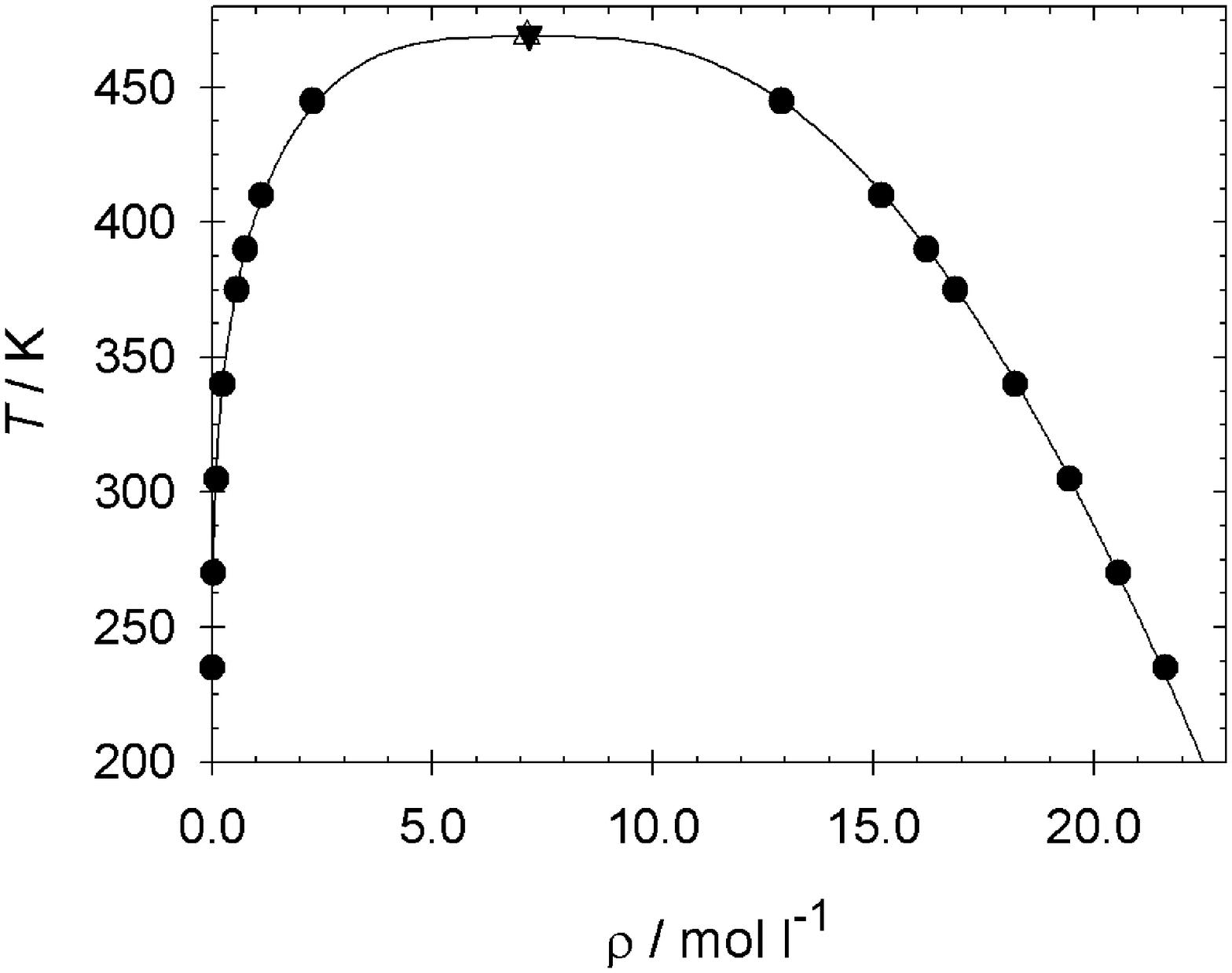}\includegraphics[scale=0.3]{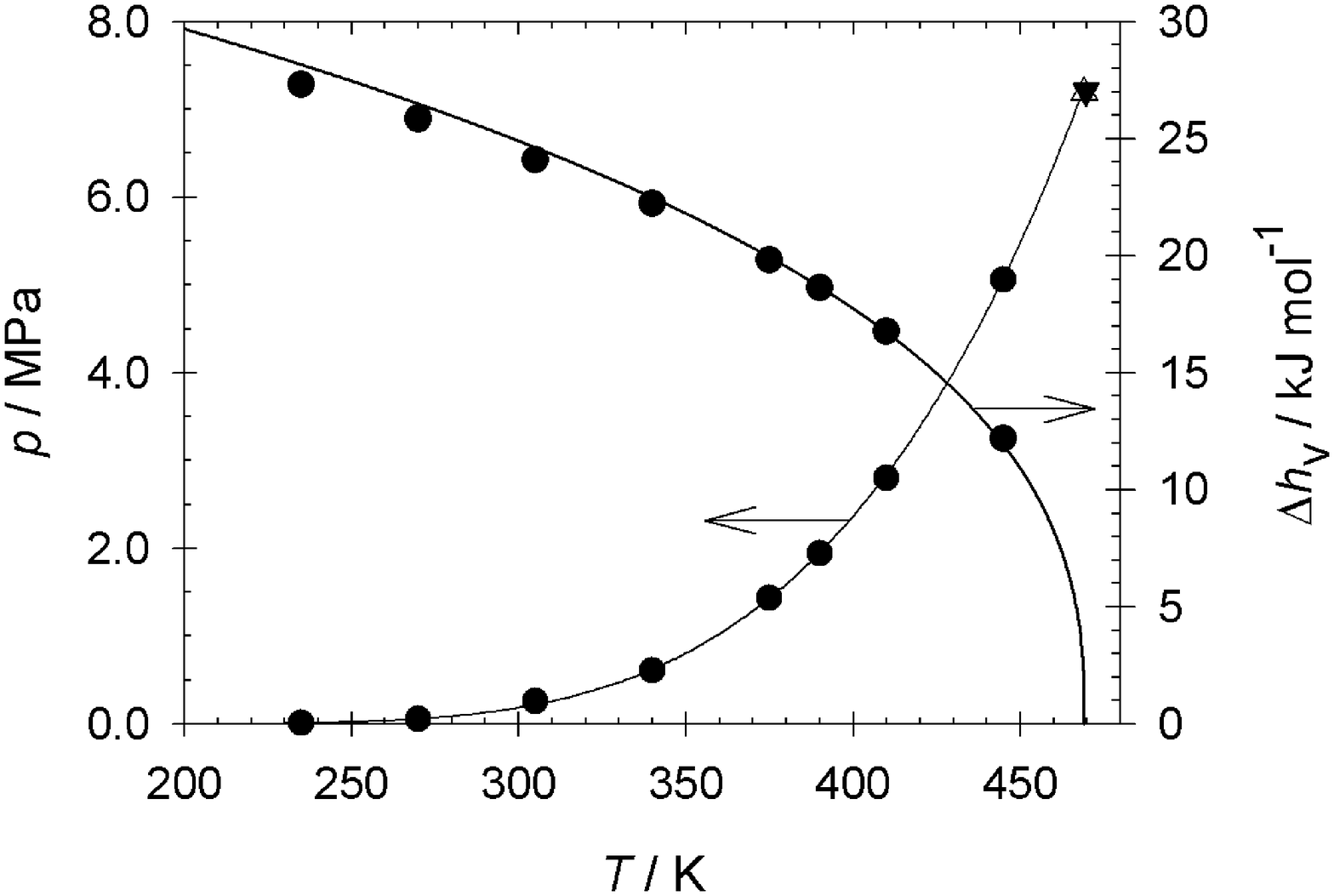}
\end{center}
\caption{Vapor-liquid equilibrium of ethylene oxide. Saturated densities (left) and vapor pressure and enthalpy of vaporization (right): {\Large $\bullet$}~simulation results, {---}~experimental data \cite{DIPPR}.\label{fig_eox_vle}}
\end{figure}

Following the optimization procedure of Stoll \cite{Stoll05}, an optimized parameter set was obtained.
The model parameters are listed in Table~\ref{tab_eox_model}. 
In Figure~\ref{fig_eox_vle}, saturated densities, vapor pressure, and enthalpy of vaporization are shown. The present model describes the vapor pressure $p_\sigma$, the saturated liquid density $\rho'$, and the enthalpy of vaporization $\Delta h_{\mathrm v}$ with mean relative deviations in the complete VLE range from triple point to critical point of $\delta p_\sigma=1.5~\%$, $\delta \rho'=0.4~\%$, and $\delta \Delta h_{\mathrm v}=1.8~\%$, respectively.


\begin{table}[htb]
\noindent
\caption{Coordinates and parameters of the present molecular model for ethylene oxide. The three Lennard-Jones sites are denoted by the molecular group which they represent, while the single dipolar site is denoted "Dipole". All coordinates are in principal axes with respect to the center of mass. The orientation of the electrostatic site is definded in standard Euler angles, where $\varphi$ is the azimuthal angle with respect to the $x$-$y$-plane and $\theta$ is the inclination angle with respect to the $z$-axis.}
\label{tab_eox_model}

\medskip
\begin{center}
\begin{tabular}{lcccccccc} \hline\hline
Interaction & $x$           & $y$   & $z$           & $\sigma$ & $\varepsilon/k_{\mathrm{B}}$ & $\theta$  & $\varphi$ & $\mu$ \\
Site        & \r{A}         & \r{A} & \r{A}         & \r{A}    & K             & $\deg$    & $\deg$    & D     \\ \hline
CH$_2$(1)   & \.0.78000     & 0     &  -0.48431     & 3.5266   & 84.739        & ---       & ---       & ---   \\
CH$_2$(2)   &  -0.78000     & 0     &  -0.48431     & 3.5266   & 84.739        & ---       & ---       & ---   \\
O(3)        & \.0\.\0\0\0\0\0 & 0     & \.0.73569     & 3.0929   & 62.126        & ---       & ---       & ---   \\
Dipole      & \.0\.\0\0\0\0\0 & 0     & \.0\.\0\0\0\0\0 & ---      & ---           & 0       & 0       & 2.459 \\ \hline\hline
\end{tabular}
\end{center}
\end{table}

The model, as specified in Table~\ref{tab_eox_model}, was used for calculating other properties discussed in the subsequent section. No further parameter adjustments were made and thus all results for properties except $p_\sigma$, $\rho'$, and $\Delta h_{\mathrm v}$ are fully predictive.

\subsection{Results and comparison to experimental data}\label{Vergleich}
Individual VLE simulation results from the present model between $235$ and $445$~K deviate by less than $3~\%$ in vapor pressure and heat of vaporization, and less than $0.5~\%$ in saturated liquid density. 

Saturated densities, vapor pressure, and enthalpy of vaporization of ethylene oxide at $375$~K are all reproduced by the present model with deviations below $1~\%$ and therefore within the experimental error given in \cite{DIPPR}.

Critical properties are in excellent agreement with experimental data, being within $1.3~\%$. The normal boiling temperature deviates by less than $0.5~\%$ from experimental data for the present model.

The second virial coefficient was calculated by direct evaluation of the intermolecular potential. It is underpredicted from simulation by $3.4~\%$.

Experimental data on isobaric heat capacity and isothermal compressibility for the saturated fluid states at 375~K have not been available to us prior to the deadline of the Simulation Challenge. Later, the IFPSC published reference values, which are used for comparison here. In the saturated liquid, the simulation overpredicts isobaric heat capacity by $13~\%$, while in the saturated vapor, it is overpredicted by $18~\%$. Reference values for isothermal compressibility are obtained from the Brelvi-O'Connell correlation \cite{BO72} for the saturated liquid and from the virial equation for the saturated vapor. Simulations for the present model yield a $39~\%$ higher result for the liquid state and a $5.6~\%$ higher result for the vapor state.

While the deviation from the reference value for the isothermal compressibility in the saturated vapor state is within the assumed uncertainty of $9.2~\%$, the one in saturated liquid state is significantly above the assumed uncertainty of $23.1~\%$. This is astonishing as the densities predicted by the present model along the bubble line are in good agreement with experimental ethylene oxide data. An additional check on our value after the deadline of the Simulation Challenge, evaluating simulation runs at different pressures for 375~K, yielded a value within the statistical uncertainty of the previous simulation result. On the other side, the Brelvi-O'Connell correlation is known to give systematic deviations to too low compressibilities for polar substances similar to ethylene oxide, e.g. sulfur dioxide or acetone, when using the critical density for reduction \cite{BO72}. Therefore, the authors believe that molecular modeling and simulation yields more reliable results here than the standard method does.

Experimental surface tension data are available between 200 and 296~K and may be extrapolated to higher temperatures using a correlation taken from \cite{DIPPR}. The simulation results give deviations of $-17~\%$.

Experimental transport properties for ethylene oxide are scarce. In fact, for viscosity and thermal conductivity at $375$~K they are only available in the saturated vapor state. Here, the results from the present model agree with the experimental values within their statistical uncertainty. After the deadline of the Simulation Challenge, the IFPSC has given reference values for viscosity in the saturated liquid state from an extrapolation of experimental data from 223 to 282~K and for thermal conductivity from the Missenard method \cite{Missenard65}. Simulation results agree with the reference values within their statistical uncertainties and the uncertainties of the reference values proposed by IFPSC.

It can be seen that the molecular model for ethylene oxide is capable to reasonably predict the wide variety of properties. This underlines that molecular modeling and simulation in combination with the chosen molecular modeling route, using experimental bubble density, vapor pressure, and enthalpy of vaporization only, can be followed for predicting properties where experimental data is insufficient.


\section{An optimized molecular model for ammonia}\label{sec_nh3}
Ammonia is a well-known chemical intermediate, mostly used in fertilizer industries; another important application is its use as a refrigerant. Due to its simple symmetric structure and its strong intermolecular interactions it is also of high academic interest both experimentally and theoretically.

In the present work a new molecular model for ammonia is proposed, which is based on the work of Krist\'of \textit{et al.} \cite{KVLRM99} and improved by including data on geometry and electrostatics from \emph{ab initio} quantum mechanical calculations.

\subsection{Selection of model type and parameterization}
For the present molecular model for ammonia, a single Lennard-Jones potential was assumed to describe the dispersive and repulsive interactions. The electrostatic interactions as well as hydrogen bonding were modeled by a total of four partial charges. This modeling approach was also followed by 
Krist\'of \textit{et al.} \cite{KVLRM99} for ammonia.

The Lennard-Jones site and the partial charges were placed according to the nucleus positions obtained from a quantum mechanical geometry optimization. For an initial model, the magnitude of the charges were taken directly from quantum mechanics. This initial model was subsequently optimized to experimental VLE data. The optimized parameters are given in Table~\ref{tab_nh3_model}.

\begin{table}[ht]
\noindent
\caption{Parameters of the molecular model for ammonia ($e = 1.6021 \cdot 10^{-19}$~C is the elementary charge).}
\label{tab_nh3_model}

\medskip
\begin{center}
\begin{tabular}{lcccccc} \hline\hline
Interaction & $x$         & $y$         & $z$         & $\sigma$ & $\varepsilon/k_\mathrm{B}$ & $q$ \\
Site        & \r{A}       & \r{A}       & \r{A}       & \r{A}    & K             & $e$      \\ \hline
N           & \.0\.\0\0\0\0 & \.0\.\0\0\0\0 & \.0.0757    & 3.376    & 182.9         & -0.9993  \\
H(1)        & \.0.9347    & \.0\.\0\0\0\0 & -0.3164     & ---      & ---           & \.0.3331 \\
H(2)        & -0.4673     & \.0.8095    & -0.3164     & ---      & ---           & \.0.3331 \\
H(3)        & -0.4673     & -0.8095     & -0.3164     & ---      & ---           & \.0.3331 \\ \hline\hline
\end{tabular}
\end{center}
\end{table}

To optimize the molecular model, the two Lennard-Jones parameters were adjusted to experimental saturated liquid density, vapor pressure, and enthalpy of vaporization using a Newton scheme as proposed by Stoll \cite{Stoll05}. These properties were chosen for the adjustment as they all represent major characteristics of the fluid region. Furthermore, they are relatively easy to be measured and are available for many components of technical interest.

\subsection{Results and discussion}
\subsubsection{Vapor-liquid equilibria}
\begin{figure}[b]
\begin{center}
\hspace*{-0.8cm}\includegraphics[scale=0.3]{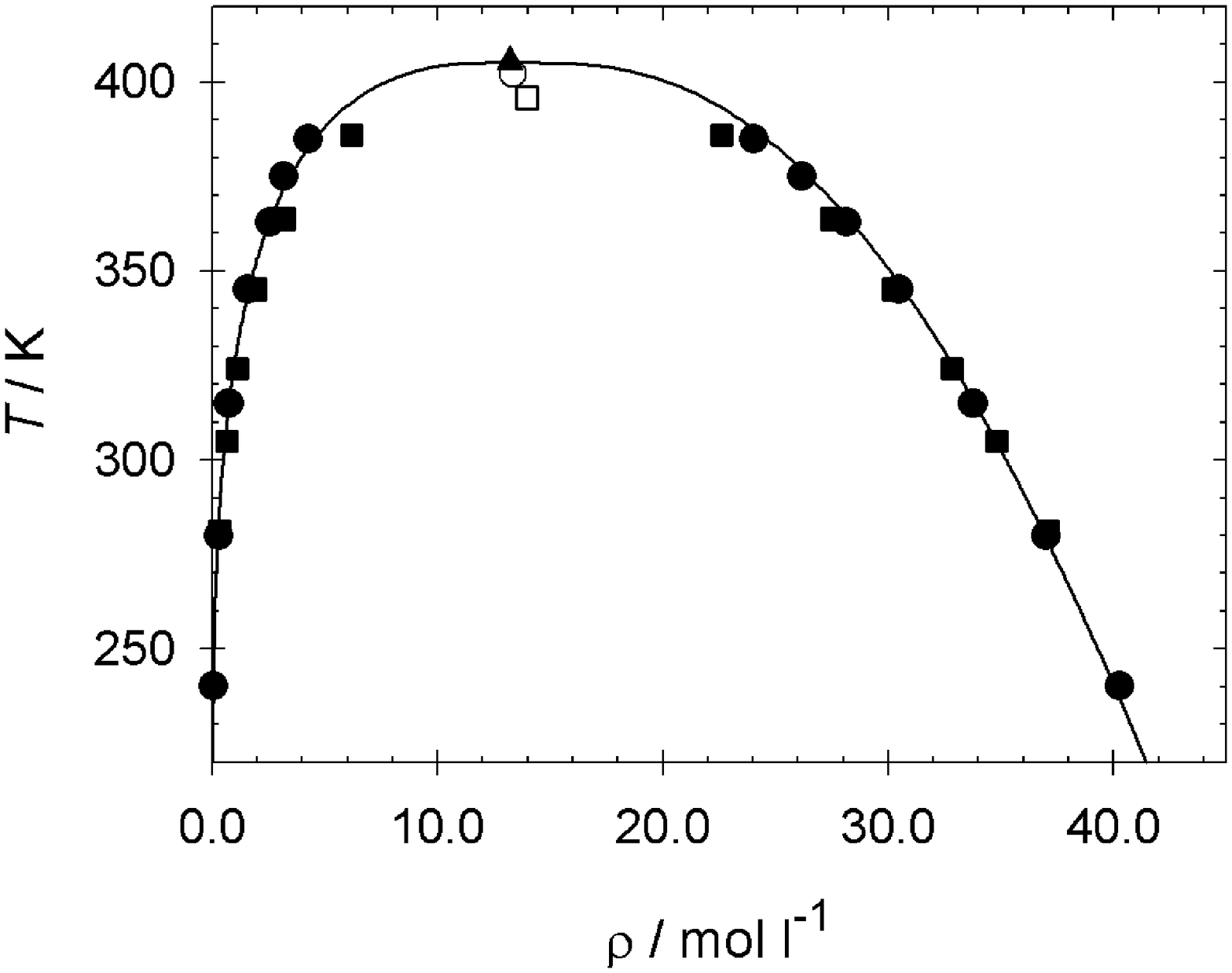}\includegraphics[scale=0.3]{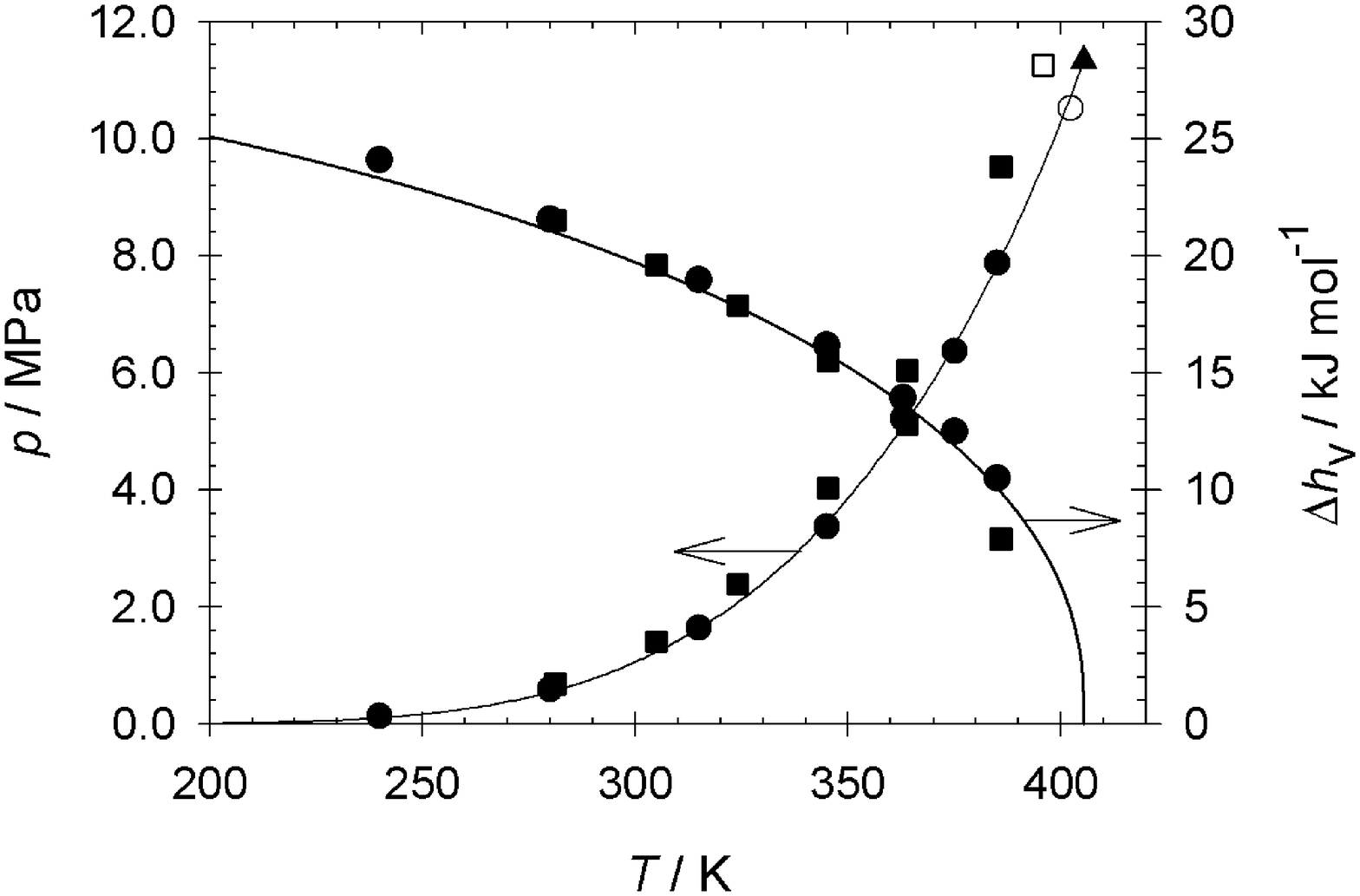}
\end{center}
\caption{Vapor-liquid equilibrium of ammonia. Saturated densities (left) and vapor pressure and enthalpy of vaporization (left): {\Large $\bullet$}~simulation results from the new molecular model, {\small $\blacksquare$}~simulation results from the model by Krist\'of \textit{et al.} \cite{KVLRM99}, {---}~experimental data \cite{DIPPR}.\label{fig_nh3_vle}}
\end{figure}

VLE results for the new model are compared to data obtained from a reference quality equation of state (EOS) \cite{THB93} in Figure~\ref{fig_nh3_vle}. This figure also includes the results that we calculated using the model from Krist\'of \textit{et al.} \cite{KVLRM99}. 

The model of Krist\'of \textit{et al.} shows noticeable deviations from experimental data. The mean unsigned errors over the range of VLE are 1.9~\% in saturated liquid density, 13~\% in vapor pressure and 5.1~\% in enthalpy of vaporization. With the new model, a significant improvement was achieved compared to the model from Krist\'of \textit{et al.} The description of the experimental VLE is very good, the mean unsigned deviations in saturated liquid density, vapor pressure and enthalpy of vaporization are 0.7, 1.6, and 2.7~\%, respectively. 

Mathews \cite{Mathews72} gives experimental critical values of temperature, density and pressure for ammonia: $T_\mathrm{c}=$405.65~K, $\rho_\mathrm{c}=$13.8~mol/l, and $p_\mathrm{c}=$11.28~MPa. Following the procedure suggested by Lotfi \textit{et al.} \cite{LVF92} the critical properties $T_\mathrm{c}=$395.82~K, $\rho_\mathrm{c}=$14.0~mol/l, and $p_\mathrm{c}=$11.26~MPa for the model of Krist\'of \textit{et al.} were calculated, where the critical temperature is underestimated by 2.4~\%. For the new model $T_\mathrm{c}=$402.21~K, $\rho_\mathrm{c}=$13.4~mol/l, and $p_\mathrm{c}=$10.52~MPa were obtained. The new model gives very good results for the critical temperature, while it underpredicts the critical pressure slightly.

\subsubsection{Homogeneous region}
In many technical applications thermodynamic properties in the homogeneous fluid region are needed. Thus, the new molecular model for ammonia was tested on its predictive capabilities for such states.

\begin{figure}[p]
\begin{center}
\hspace*{-0.2cm}\includegraphics[scale=0.3]{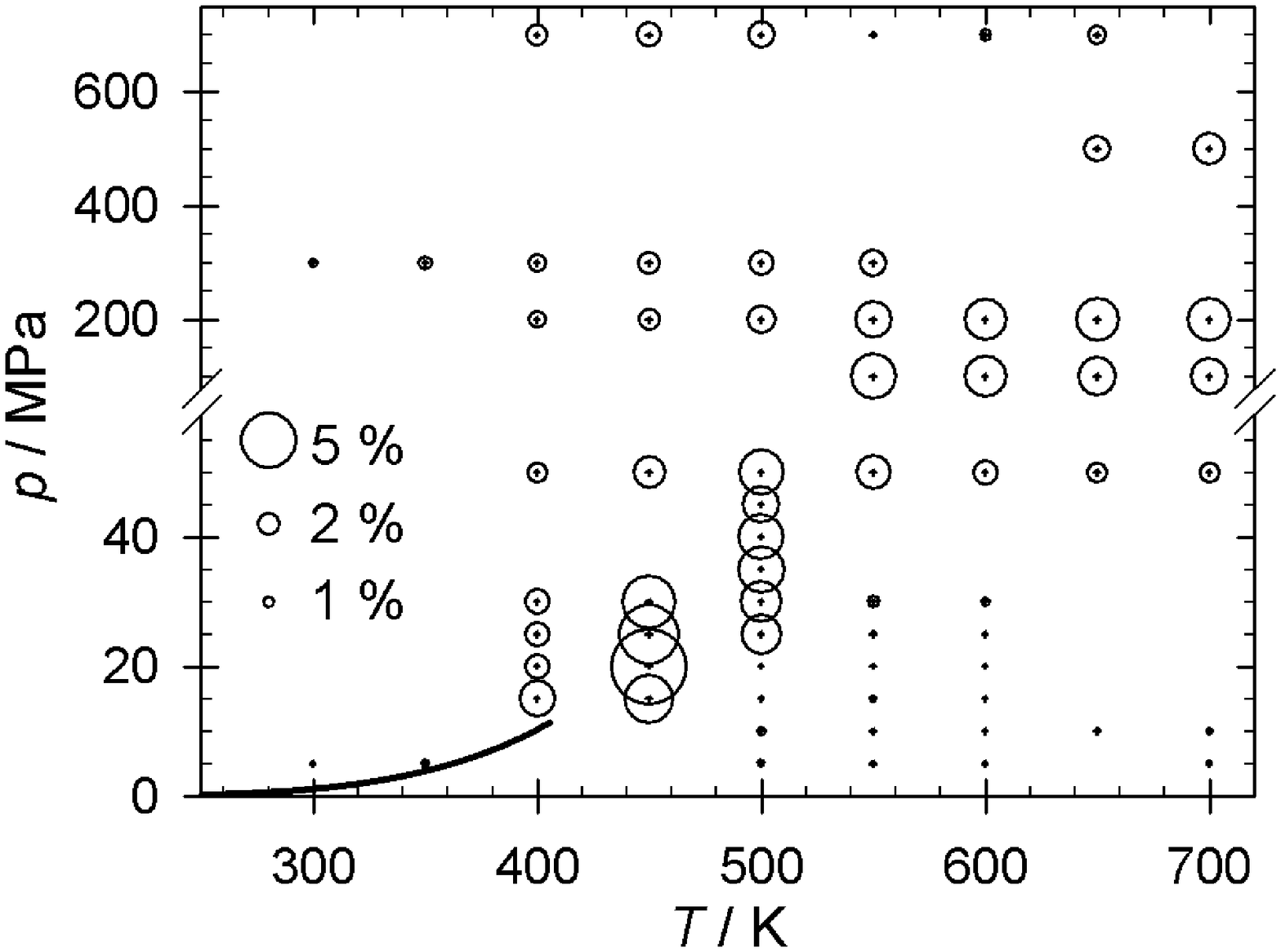}\includegraphics[scale=0.3]{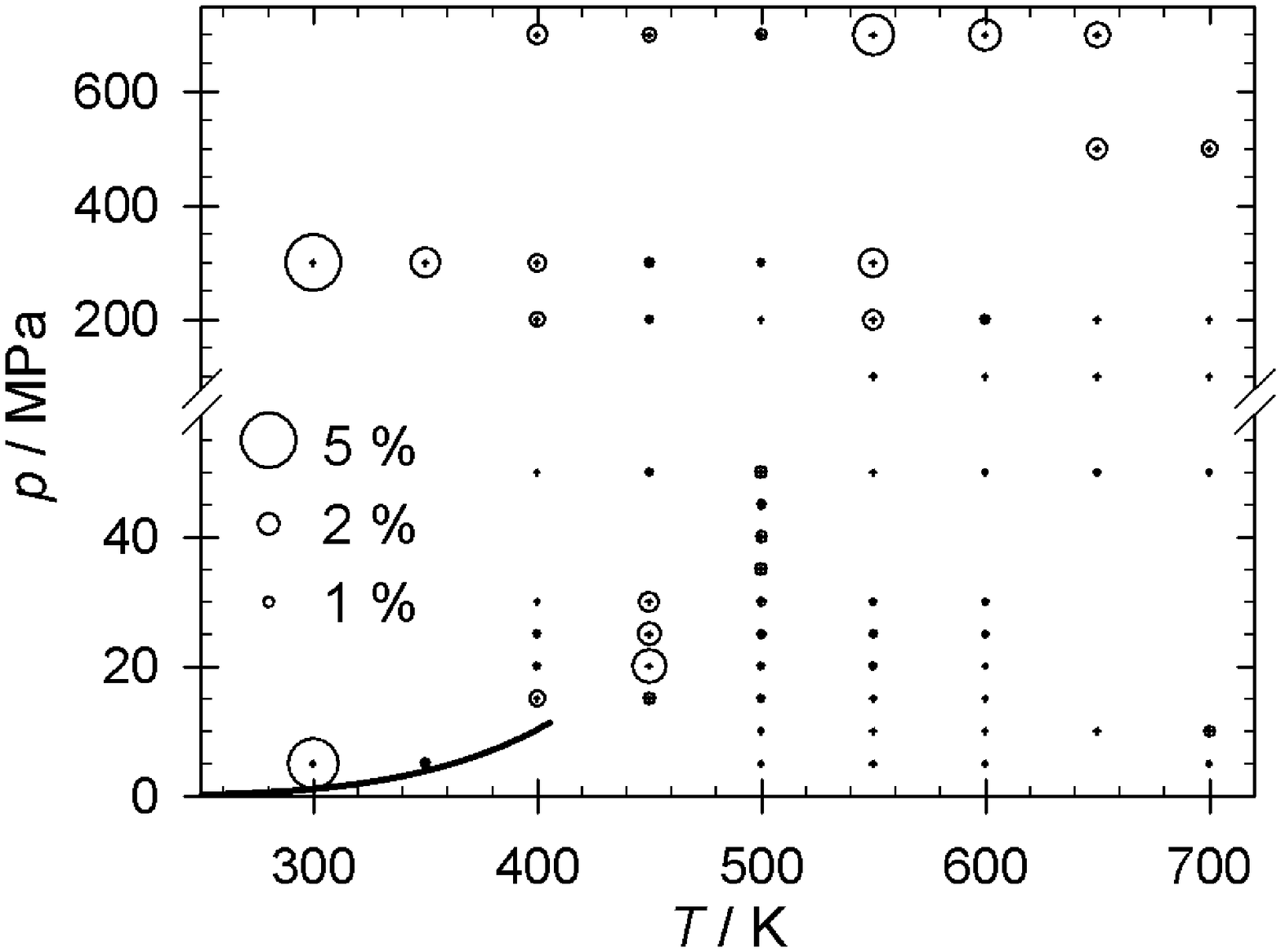}
\end{center}
\caption{Relative deviations for the density (left) and the enthalpy (right) between simulation and reference EOS \cite{THB93} ($\delta z = (z_{\mathrm{sim}} - z_{\mathrm{eos}}) / z_{\mathrm{eos}}$) in the homogeneous region: {\Large $\circ$}~simulation data of new model, {---}~vapor pressure curve. The size of the bubbles denotes the relative deviation as indicated in the plot.\label{fig_nh3_hom}}
\end{figure}

Thermal and caloric properties were predicted with the new model in the homogenous liquid, vapor and supercritical fluid region. In total, 70 state points were regarded, covering a large range of states with temperatures of up to 700~K and pressures of up to 700~MPa. In Figure~\ref{fig_nh3_hom}, relative deviations between simulation and reference EOS \cite{THB93} in terms of density and enthalpy are shown. The deviations are typically below 3~\% for density with the exception of the extended critical region, where a maximum deviation of 6.8~\% is found, and below 5~\% for enthalpy.

These results confirm the modeling procedure. By adjustment to VLE data only, quantitatively correct predictions in most of the technically important fluid region can be obtained.

%
%

\subsubsection{Structural quantities}
\begin{figure}[p]
\begin{center}
\includegraphics[scale=0.3]{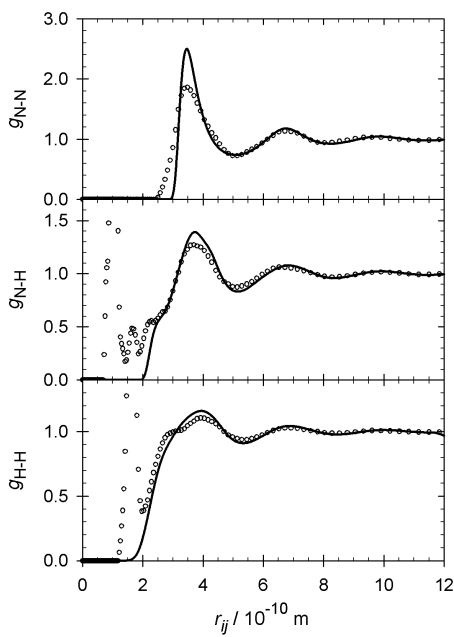}
\end{center}
\caption{Pair correlation functions of ammonia: {---}~simulation data with the new model, {\small $\circ$}~experimental data \cite{RNRAS95}.\label{fig_nh3_rdf}}
\end{figure}

Due to its scientific and technical importance, experimental data on the microscopic structure of liquid ammonia are available. Ricci \textit{et al.} \cite{RNRAS95} applied neutron diffraction and published all three types of atom-atom pair correlations, namely nitrogen-nitrogen (N-N), nitrogen-hydrogen (N-H), and hydro\-gen-\-hydro\-gen (H-H). In Figure~\ref{fig_nh3_rdf}, these experimental radial distribution functions for liquid ammonia at 273.15~K and 0.483~MPa are compared to present predictive simulation data based on the new ammonia model.

It was found that the structural properties are in very good agreement, although no adjustment was done regarding these properties. The atom-atom distance of the first three layers is predicted correctly, while only minor overshootings in the first peak are found. Please note that the first experimental peak in $g_\mathrm{N-H}$ and $g_\mathrm{H-H}$ show intramolecular pair correlations, which are not included in the present model.

In the experimental radial distribution function $g_\mathrm{N-H}$ the hydrogen bonding of ammonia can be seen at 2--2.5~\r{A}. Due to the simplified approximation by eccentric partial charges, the molecular model is not capable to describe this effect completely. But even with this simple model a small shoulder at 2.5~\r{A} is obtained.

\section{Nucleation processes during condensation}\label{sec_nuc}
Understanding homogeneous nucleation is required to develop an accurate theoretical approach to nucleation that extends to more complex and technically more
relevant heterogeneous systems. MD simulations can be used to investigate the condensation of homogeneous vapors at high supersaturations.

The nucleation rate $J$ is influenced to a large extent by the surface energy of droplets, which also determines how many droplets are formed and from which size on they become stable. The accuracy of different theoretical expressions for the surface energy can be assessed by comparison to results from MD simulations.

\subsection{Nucleation theory}
Classical nucleation theory (CNT) was developed by Volmer and Weber \cite{VW26} in the 1920s and further extended through many contributions over the following decades \cite{FRLP66}. It is founded on the capillarity approximation: droplets emerging during nucleation are assumed to have the same thermodynamic properties as the bulk saturated liquid. In particular, the specific surface energy of the emerging nano-scaled droplets is assumed to be the surface tension of the planar phase boundary in equilibrium.

Laaksonen, Ford, and Kulmala \cite{LFK94} (LFK) proposed a surface energy coefficient that depends on the number of molecules in the droplet, i.e. the size of the droplet. Tanaka \textit{et al.} \cite{TKTN05} found that this expression leads to nucleation rates which agree with their simulation results. It was shown both theoretically \cite{BB99, KB49} and by simulation \cite{VKFH06, NL07} that the surface tension acting in the curved interface of nano-scaled droplets is actually much lower than in a planar interface. This finding underlines the necessity of a correction to CNT.


\subsection{Simulation method}
Both the critical droplet size and the nucleation rate can be determined by molecular simulation. After an initial period of equilibration, the steady state distribution of droplets is established. The critical droplet size, i.e. the size from which on droplets more probably grow than decay, can either be calculated from the equilibrium distribution \cite{YM98} or by the "nucleation theorem" \cite{OK94}. Both methods require data on the nucleation rate $J$, i.e. the number of macroscopic droplets emerging per volume and time in a steady state at constant supersaturation. From an MD simulation of a supersaturated vapor, this rate can straightforwardly be extracted by counting the droplets that exceed a certain threshold size. This method was proposed by Yasuoka and Matsumoto \cite{YM98}, who found that as long as this threshold is larger than the critical droplet size, its precise choice hardly affects the observed value of $J$.


The nucleation rate $J$ is the number of nuclei formed per volume and time in a supersaturated vapor
\begin{equation}
J(\iota) = \sum_{n > \iota} \frac{\partial \rho_n}{\partial t}~,
\end{equation}
for a threshold size $\iota$ larger than the critical size, measured after an initial temporal delay required to form sufficiently large nuclei. For nucleation rates obtained from molecular simulations, a very low size threshold is used, roughly of the same order of magnitude as the critical size. Because the choice of the threshold may influence the result, it must be indicated explicitly, i.e. as $J(\iota)$ instead of $J$.

For simulations in the canonical ensemble, it has to be taken into account that as the condensation proceeds, the density of the remaining vapor decreases and the pressure in the vapor is reduced significantly. This causes larger nuclei to be formed at a lower rate. Thus, the nucleation rates are given together with pressure values which were taken in the middle of the interval where the value of $J(\iota)$ was obtained by linear approximation.

\begin{figure}[b]
\begin{center}
\includegraphics[scale=0.23]{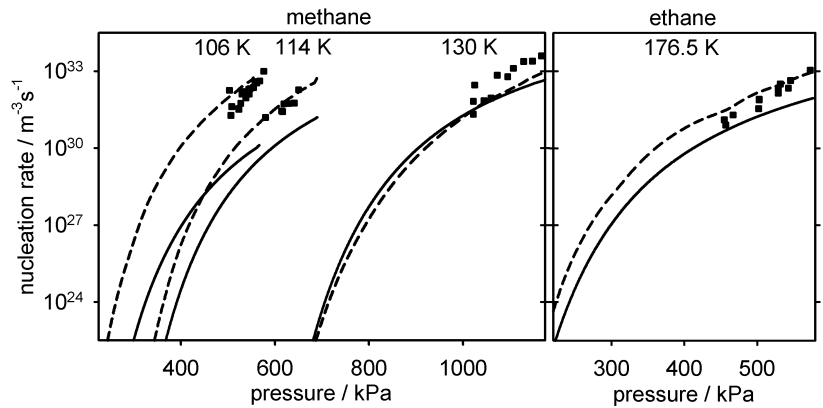}
\end{center}
\caption{Nucleation rates of methane and ethane at low temperatures from simulation according to the Yasuoka-Matsumoto method and different threshold sizes ({\small $\blacksquare$}) as well as CNT (---) and LFK (-~-~-).\label{fig_nuc_1}}
\end{figure}

For the present study, methane was modeled as a LJ fluid and ethane as a rigid two-center LJ fluid with an embedded point quadrupole \cite{VSH01}. Furthermore, the truncated and shifted LJ (LJTS) fluid with a cutoff radius of 2.5 times the size parameter $\sigma$ was studied. The LJTS fluid is an accurate model for fluid noble gases and methane \cite{VKFH06}.

Within a typical computation time of about 24 hours, a time interval of a few nanoseconds can be simulated for systems with a volume of 10$^{-21}$ m$^{3}$. Given that the nucleation rate is determined as units per volume and time, only values that exceed 10$^{30}$ m$^{-3}$s$^{-1}$ can be obtained from molecular simulation. However, only nucleation rates below 10$^{23}$ m$^{-3}$s$^{-1}$ can be measured in experiments at present. Therefore, computational power is crucial to keep the gap between simulation and experiment as small as possible.

In Figure~\ref{fig_nuc_1} the nucleation rates from MD simulation are shown in comparison to predictions from CNT and LFK theories. At the given state points, simulation results confirm both CNT and LFK, deviations are throughout lower than three orders of magnitude.

\subsection{Simulation results}

\begin{figure}[t]
\begin{center}
\includegraphics[scale=0.33]{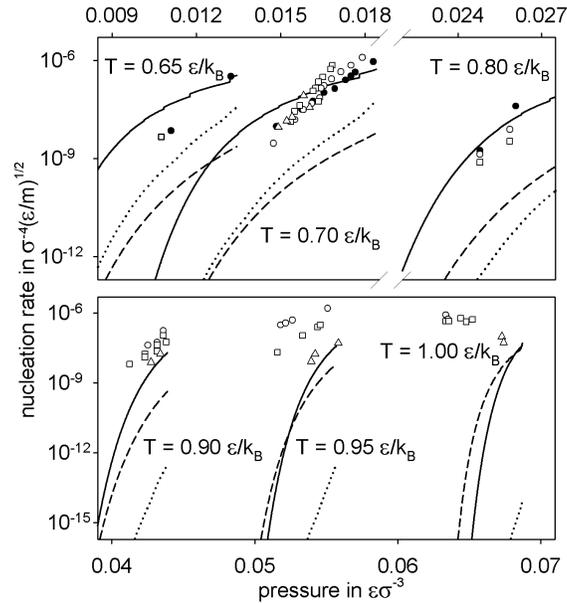}
\end{center}
\caption{Nucleation rate of the LJTS fluid over supersaturated pressure from the present direct simulations for different threshold values ({\Large $\bullet$}~$i = 25$, {\Large $\circ$}~$i = 50$, {\small $\blacksquare$}~$i \in \{75, 100\}$, $\triangle$~$i \geq 150$) and following the standard CNT (-~-~-), the LFK modification of CNT ($\cdots$), and the new SPC modification of CNT (---).\label{fig_nuc_2}}
\end{figure}

However, from a series of simulations it was seen that standard CNT only predicts the nucleation rate $J$ with an acceptable accuracy, but leads to deviations for the critical droplet size; the LFK modification provides excellent predictions for the critical size but not for the temperature dependence of $J$. LFK underpredicts the nucleation rate at high temperatures by several orders of magnitude. Both theories assume an inappropriate curvature dependence of the surface tension, although for this essential property of inhomogeneous systems a qualitatively correct expression is known since the 1940s \cite{Tolman49}. With the collected simulation data on critical droplet size and nucleation rate over a broad range of temperatures, enough quantitative information is available to formulate a more adequate modification of CNT. For the LJTS fluid, a surface property corrected (SPC) modification of CNT was developed which postulates a non-spherical surface and is based on surface tension values obtained from molecular simulation. As shown in Figure~\ref{fig_nuc_2}, the SPC modification accurately describes the nucleation rate of the LJTS fluid.

%
%

\section{Computing performance}

\begin{table}[b!]
\noindent
\caption{Scaling of the massively parallel program $ls1$ on HP~XC4000. Particle numbers $N$ are 0.5--2 million.}
\label{tab_ls1_scaling}

\medskip
\begin{center}
\begin{tabular}{cc} \hline\hline
CPUs & CPU time (per time step and molecule) \\ \hline
 24  &  0.022 ms \\
 32  &  0.021 ms \\
 40  &  0.021 ms \\
100  &  0.025 ms \\
125  &  0.023 ms \\ \hline\hline
\end{tabular}
\end{center}
\end{table}

All simulations presented in Sections~\ref{sec_eox} and \ref{sec_nh3} were carried out with the MPI based molecular simulation program $ms2$ developed in our group. The parallelization of the molecular dynamics part of $ms2$ is based on Plimptons force decomposition algorithm \cite{Plimpton94}. The nucletion simulations were carried out with the massivly parallel program $ls1$. Here, the parallelization relies on spatial domain decomposition for parallel simulations \cite{BV05} due to the large size of the simulated systems (up to 2 million particles).

With $ms2$ typical simulation runs to determine the vapor-liquid equilibrium employ 4--8 CPUs running for 4--6 hours. For model optimization or the comprehensive study on mixtures, a large number of independent simulations are necessary and can be performed in parallel. For the prediction of other thermophysical properties, additional simulations were performed using up to 32 CPUs and running up to 72 hours depending on the system size and desired property.

For the simulation of nucleation processes, very large systems must be considered. Here, the massively parallel program $ls1$ was used. Table~\ref{tab_ls1_scaling} demonstrates the good scaling of the program on the HP~XC4000 cluster at the Steinbuch Center for Computing, Karlsruhe.






\end{document}